\documentclass[preprint]{emulateapj}

\shorttitle{Adaptive Optics Discovery of a SN 2004ip}
\shortauthors{Mattila et al.}

\def\kms{\ifmmode{\rm km\,s^{-1}}\else\hbox{$\rm km\,s^{-1}$}\fi}

\begin{document}


\title{Adaptive Optics Discovery of Supernova 2004ip in the Nuclear Regions of the Luminous Infrared Galaxy IRAS 18293-3413}


\author{S. Mattila\altaffilmark{1}, P. V\"ais\"anen\altaffilmark{2}, D. Farrah\altaffilmark{3}, 
A. Efstathiou\altaffilmark{4}, W.P.S. Meikle\altaffilmark{5}, T. Dahlen\altaffilmark{6}, 
C. Fransson\altaffilmark{7}, P. Lira\altaffilmark{8}, P. Lundqvist\altaffilmark{7}, 
G. \"Ostlin\altaffilmark{7}, S. Ryder\altaffilmark{9}, J. Sollerman\altaffilmark{10}}
\altaffiltext{1}{Astrophysics Research Centre, School of Mathematics and Physics, Queen's
University Belfast, BT7 1NN, UK; s.mattila@qub.ac.uk}
\altaffiltext{2}{South African Astronomical Observatory, PO Box 9, Observatory 7935, South Africa.}
\altaffiltext{3}{Department of Astronomy, Cornell University, 106 Space Sciences, Ithaca, NY 14853.}
\altaffiltext{4}{School of Computer Science and Engineering, Cyprus College, 6 Diogenes Street, Engomi, 1516 Nicosia, Cyprus.}
\altaffiltext{5}{Astrophysics Group, Blackett Laboratory, Imperial
College London, Prince Consort Road, London SW7 2AZ, UK.}
\altaffiltext{6}{Space Telescope Science Institute, 3700 San Martin Drive, Baltimore, MD 21218.}
\altaffiltext{7}{Department of Astronomy, Stockholm University, AlbaNova, SE-106 91 Stockholm, Sweden.}
\altaffiltext{8}{Departamento de Astronomía, Universidad de Chile, Santiago, Chile.}
\altaffiltext{9}{Anglo-Australian Observatory, PO Box 296, Epping, NSW 1710, Australia.}
\altaffiltext{10}{Dark Cosmology Centre, Niels Bohr
Institute, University of Copenhagen, Juliane Maries Vej 30, DK-2100
Copenhagen, Denmark.}




\begin{abstract}
We report a supernova discovery in K$_{\rm S}$-band images from the NAOS CONICA adaptive optics (AO)
system on the ESO Very Large Telescope (VLT). The images were obtained as part of a near-infrared search
for highly-obscured supernovae in the nuclear regions of luminous and ultraluminous infrared galaxies.
\object{SN 2004ip} is located within a circumnuclear starburst at 1.4 arcsec (or 500 pc) projected distance from the K-band
nucleus of the luminous infrared galaxy \object{IRAS 18293-3413}. The supernova luminosity and light curve are consistent with
a core-collapse event suffering from a host galaxy extinction of up to about 40 magnitudes in V-band which is as
expected for a circumnuclear starburst environment. This is the first supernova to be discovered making use of
AO correction and demonstrates the potential of the current 8-meter class telescopes equipped with AO in
discovering supernovae from the innermost nuclear regions of luminous and ultraluminous infrared galaxies.
\end{abstract}

\section{Introduction}
The observed rate at which stars more massive than $\sim$8 M$_{\odot}$~explode as core-collapse
supernovae (CCSNe) can be used as a direct measure of the current
star formation rate (SFR), given a reasonable assumption for the initial mass function (IMF). In fact,
CCSNe are beginning to be used as probes of the SFR at both low and high redshift ({\it e.g.} Cappellaro et al.
1999; Dahlen et al. 2004; Cappellaro et al. 2005) with the aim of providing a new independent measurement of the star formation
history of the Universe. A large fraction of the massive star formation at high-$z$ took place in luminous (L$_{{\rm IR}}$ $>$ 10$^{11}$ L$_{\odot}$)
and ultraluminous (L$_{{\rm IR}}$ $>$ 10$^{12}$ L$_{\odot}$) infrared (IR) galaxies (henceforth LIRGs and ULIRGs)
and their high SFRs can be expected to result in CCSN rates a couple of orders of magnitude higher than in ordinary field galaxies.
However, most of the SNe occurring in LIRGs and ULIRGs are likely to be obscured by large amounts of dust in the nuclear starburst
environment and have therefore, even in the local Universe, remained undiscovered by optical SN searches.
The existence of hidden SN factories in the nuclei of LIRGs and ULIRGs has already been demonstrated by high-resolution radio
observations. For example, VLBI observations of the nearby ULIRG, Arp~220, have revealed luminous radio SNe within its innermost
$\sim$150 pc nuclear regions at a rate indicating a SFR high enough to power its entire IR luminosity (Lonsdale et al. 2006; Parra et al. 2006).
However, not all CCSNe are likely to become luminous at radio wavelengths, and therefore radio searches for SNe can
only provide a lower limit for the real SN rates.

In the near-IR K-band the extinction is strongly reduced while the sensitivity
of ground-based observations is still good, making searches for SNe in starburst galaxies, LIRGs and ULIRGs more promising (Van Buren et al. 1994a;
Grossan et al. 1999; Mattila \& Meikle 2001 henceforth MM01; Maiolino et al. 2002; Mannucci et al. 2003; Mattila et al. 2004). Furthermore,
the recent introduction of adaptive optics (AO) for 8-meter class telescopes now enables near-IR searches for SNe at a spatial resolution
comparable to some of the radio studies. In this letter we report a SN discovery in the circumnuclear starburst of \object{IRAS 18293-3413} which is a LIRG at a redshift
of $z$ = 0.0182 (Strauss et al. 1992) corresponding to a distance of 79 Mpc (H$_{0}$ = 70 km~s$^{-1}$Mpc$^{-1}$, $\Omega_{\Lambda}$=0.7 and $\Omega_{\rm M}$=0.3).
The SN was detected as a result of our K$_{\rm S}$-band search of highly obscured CCSNe in a sample of 15 nearby LIRGs and ULIRGs using the NAOS CONICA (NACO) 
AO system on the ESO Very Large Telescope (VLT).

\setcounter{footnote}{0}
\section{Observations and Results}

\subsection{NACO observations and data reduction}
\object{IRAS 18293-3413} was observed for the first time with VLT/NACO (Rousset et al. 2003; Lenzen et al. 2003) in service mode on 2004 May 4.3 UT
in the K$_{\rm S}$-band with the S27 camera (0.027 arcsec/pixel) using the autojitter imaging sequence
(see Table~1 for details on the observing parameters and conditions). The AO correction was performed
using the visual wavefront sensor (VIS-WFS) with a natural guide star of m$_{\rm V}$$\sim$15, 25" northeast
from the science target. The observations were repeated on 2004 Sept. 13.1 UT  to allow SN detection by subtracting
the images from the two epochs. The data obtained on Sept. 13 were immediately reduced and searched for any obvious SNe
(for details see below and Section 2.2). This analysis produced a clear detection of a new point source in the
Sept. 13 data. We therefore triggered the Target of Opportunity (ToO) part of the programme
to obtain a deep repeat observation in the K$_{\rm S}$-band on Sept. 15.0 UT, two days after the discovery image.
About two weeks later, on Sept. 27.1 UT we triggered VLT ToO again, this time to obtain both light curve and H-K
colour information for the SN. The SN was reported on the `CBAT Possible Supernovae Page'\footnote{http://cfa-www.harvard.edu/iau/CBAT$\_$PSN.html} 
immediately after the discovery and was later assigned the designation \object{SN 2004ip} (Mattila et al. 2007).

The NACO data were reduced using the IRAF DIMSUM package and IDL. The jittered on-source frames were median combined
to form a sky frame. The sky subtracted images were de-dithered making use of the centroid coordinates of a bright
field star visible in all the frames, and the de-dithered frames were median combined excluding frames with the lowest strehl
ratios (SR) (see Table~1).

\begin{table}[t]
\caption{Log of VLT/NACO observations of IRAS 18293-3413.}
\begin{tabular}{cccccc}
Date (UT)  & Band    & Integration                &Seeing       & Airmass       & SR\\
2004       &         & (s)                        &(arcsec)     &               & (\%)\\ 
\tableline\tableline
May 4.3    & K$_{\rm S}$ & 24 $\times$ 90  & 0.47--0.79   & 1.08--1.19  & 20--30 (22)\\
Sept. 13.1 & K$_{\rm S}$ & 23 $\times$ 90  & 0.65--0.90   & 1.15--1.28  & 15--20 (21)\\
Sept. 15.0 & K$_{\rm S}$ & 63 $\times$ 90  & 0.70--1.25   & 1.01--1.15  & 5--20 (41)\\
Sept. 27.1 & H           & 68 $\times$ 90  & 0.53--1.15   & 1.11--1.70  & 5--15 (38)\\
Sept. 27.1 & K$_{\rm S}$ & 15 $\times$ 90  & 0.69--0.81   & 1.81--2.13  & 5--10 (6)\\
\tableline
\end{tabular}
\tablecomments{
The strehl ratios (SR) of frames used for the combined images are listed
in col.~6 with the number of frames in brackets.
}
\vspace{+0.7mm}
\end{table}

\begin{figure*}[t]
\epsscale{1}
\plotone{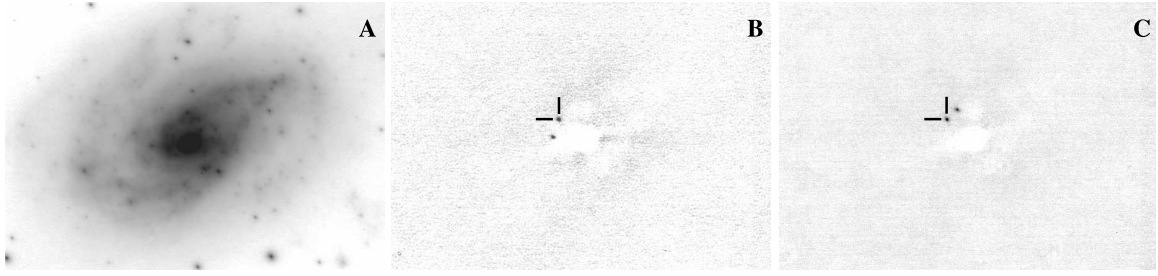}
\caption{
A 16.5"$\times$11.5" subsection of the NACO K$_{\rm S}$-band image of IRAS 18293-3413 obtained on 2004 Sept. 
15.0 UT with North up and East to the left (A).
The result of image matching and subtraction of the NACO K$_{\rm S}$-band reference image (May 4.3)
from the images obtained on Sept. 13.1 (B) and Sept. 15.0 (C). SN 2004ip (indicated by tick-marks)
is clearly detected in the subtracted images at 1.14" east and 0.78" north of (or 500 pc projected
distance) from the K-band nucleus. Simulated sources of m(K$_{\rm S}$) = 18.7 are shown southeast
and northwest of the SN in the two subtracted images, respectively. A strong negative residual
is visible at the location of the galaxy nucleus in the subtracted images. All the images are shown with an 
inverted brightness scale.}
\end{figure*}

\subsection{Supernova detection and photometry}
The images from Sept. 13, Sept. 15, and Sept. 27 were rotated and shifted to align them to the reference image
(from May 4) using bright isolated stars around the galaxy. We used a slightly modified version of the Optimal Image Subtraction method
(Alard $\&$ Lupton 1998, Alard 2000) as implemented in ISIS 2.2 to derive a convolution kernel to match the better seeing image to the
image with the poorer seeing and to match the intensities and background differences between the frames. In MM01 we
demonstrated the feasibility of this method in nuclear SN detection. The subtracted images show a clear point source (with its peak at the
$\sim$20$\sigma$ and $\sim$7$\sigma$ level in Sept. 13-15 and Sept. 27 epochs, respectively) at 1.14" east and 0.78" north from the K$_{\rm S}$-band
nucleus of the galaxy (see Fig.~1). However, without a suitable reference image for image subtraction we were unable to detect the SN in our H-band data.
Using the world coordinate system from a 2MASS image of \object{IRAS 18293-3413} we obtained R.A. = 18h32m41s.26 and Decl. = $-34^{o}11'26.7"$
(equinox 2000.0) for the SN with an estimated precision of $\pm$0.4".

We adopted the photometric zeropoints provided by ESO for our nights of observations and
estimate them to be accurate to $\pm$0.03 magnitudes in comparison to 2MASS data.
Since the SR changes as a function of the distance from the AO guide star, to measure the SN photometry
with the associated uncertainties, we used two isolated bright stars lying,
respectively, 12" and 38" from the guide star (7.1"N, 18.3"E and 5.8"S, 11.5"W of the galaxy K-band nucleus) to provide 
a point spread function (PSF) with a higher and lower SR compared to the actual SN. The two PSFs were scaled to
correspond to magnitudes ranging between 18.5 and 19.1 in steps of 0.1. Sources with each magnitude and PSF
were then simulated in the images at twenty different locations evenly distributed around the galaxy nucleus
at the same offset as the real SN. The sources were added to the images one at a time before the images were
aligned, matched and subtracted. The magnitudes for both the real and simulated SNe were measured from each
subtracted image using aperture photometry with a 0.15" aperture radius, and a sky annulus between 0.20" and
0.30". We then estimated a magnitude for the SN by comparing its photometry with the mean of the counts measured
for the 20 simulated sources with each magnitude and PSF. This yielded SN magnitudes of m(K$_{\rm S}$) =  18.74 and
18.64 for Sept. 13 using the two different PSFs, respectively. Similarly, we obtained m(K$_{\rm S}$) =  18.66 and 18.72
for Sept. 15. We adopted the standard deviation ($\sigma$) of the 20 measurements as the uncertainty in our SN photometry,
yielding $\sigma$ = 0.12 and 0.09 for the two epochs, respectively. Measuring with a 50\% larger aperture and sky annulus
gave very similar magnitudes as expected since the photometry is performed relative to the simulated sources, only the uncertainties 
were increased. During the Sept. 27 K$_{\rm S}$-band observation the airmass was high, resulting
in a low SR and therefore wider PSF for the images. For this epoch stronger residuals were also present in the subtracted
images and we therefore excluded ten of the twenty simulated sources which were affected by much stronger residuals than the SN.
For the Sept. 27 data we used an aperture radius of 0.30", and a sky annulus
between 0.40" and 0.50". This yielded m(K$_{\rm S}$) = 18.82 and m(K$_{\rm S}$) = 18.99 with $\sigma$ = 0.14 using the
two PSFs, respectively. We adopt the average of the magnitudes obtained with the two different PSFs as the best estimate for
the SN magnitude. The uncertainties were obtained as a quadrature sum of the $\sigma$ value and the error in the zeropoint.

\subsection{Supernova type and extinction}
Without a spectroscopic observation of the SN its confirmation and possible typing have to be based on
photometry and other available information. An asteroid origin for the source was immediately ruled out since the
images separated by 2 days show the SN in exactly the
same position (within $\pm$0.02"). Furthermore, the SN is still at precisely
the same location in the 3rd epoch images, obtained 2 weeks from the
discovery. Below we model the near- to far-IR SED of \object{IRAS 18293-3413}
and find a negligible contribution from an active galactic nucleus (AGN) to its luminosity. Also, the presence of an AGN
is not favoured by the X-ray results of Risaliti et al. (2000) and the fact that \object{IRAS 18293-3413}
shows an `H II galaxy' spectrum in the optical (Veilleux et al. 1995). Therefore, an AGN
flare can be ruled out as origin for the source. A variable foreground star also provides a very improbable alternative explanation
given the small field of view (42"$\times$42", {\it i.e.}, 1/(3$\times$10$^{8}$) of the whole sky)
of our combined NACO frames. For example, the likelihood of catching a foreground nova would be extremely small
given the Galactic nova rate of $\sim$35 yr$^{-1}$ (Shafter 1997). Furthermore, a classical nova in
\object{IRAS 18293-3413} would be several magnitudes fainter than the detected SN (Hachisu \& Kato 2006). We, therefore,
conclude that anything other than a SN origin for our source appears extremely unlikely.

To estimate the SFR for \object{IRAS 18293-3413} we compiled its 1-100 micron archival (2MASS and
IRAS) photometry and fitted the resulting SEDs (see Fig.~2) both with the starburst models of
Efstathiou {\it et al.} (2000), and the cirrus model of Efstathiou \& Rowan-Robinson (2003).
The combination of the starburst and cirrus model provides a very good fit to the near-
to far-IR SED and is consistent with a negligible contribution from an AGN.
The model assumes an exponentially decaying starburst with a time constant of 40 Myrs
and an age of 70 Myrs, and an underlying 12.5 Gyr old stellar population.
The cirrus A$_{\rm V}$ is assumed to be 1.7 and the initial A$_{\rm V}$ of the molecular clouds that constitute
the starburst is 100. The SFR at the peak is 285 M$_{\odot}$~yr$^{-1}$,
and 135 M$_{\odot}$~yr$^{-1}$ when averaged over the duration of the starburst.
We estimate a rest frame 8-1000 $\mu$m luminosity of 5.8 $\times$ 10$^{11}$ L$_{\odot}$
and a bolometric luminosity of 6.5 $\times$ 10$^{11}$ L$_{\odot}$ for \object{IRAS 18293-3413}.

Adopting the average SFR and assuming CCSN progenitors between 8 and 50 M$_{\odot}$~and a Salpeter IMF with
cut-offs at 0.1 and 125M$_{\odot}$~(see also MM01), we obtain a CCSN rate of $\sim$1 yr$^{-1}$
for \object{IRAS 18293-3413}. We note that this CCSN rate also agrees with the model predictions of Genzel et al.
(1998) for an exponentially decaying starburst with similar time constant, age and bolometric luminosity.
Recent observational evidence shows that the Type Ia SN (SN Ia) rate is likely to depend on both the stellar mass and
the mean SFR of their host galaxies (see Sullivan et al. 2006 and references therein). The rate of SNe Ia associated with
the old stellar population is not expected to be enhanced in \object{IRAS 18293-3413} compared with normal field galaxies but the rate associated with
the recent star formation could be higher. This rate has been found to depend on the host galaxy SFR averaged
over the last 500 Myrs (SFR$_{\rm ave}$) according to $\sim$4 $\times$ 10$^{-4}$ SNe yr$^{-1}$ / SFR$_{\rm ave}$ (Sullivan et al. 2006).
Using 135 M$_{\odot}$~yr$^{-1}$ as an upper limit for SFR$_{\rm ave}$ we obtain an upper limit for its SN Ia rate of $\sim$0.05 yr$^{-1}$
which is $\sim$20$\times$ smaller than the estimated CCSN rate. We therefore conclude that SN 2004ip was very
likely a core-collapse event.

\begin{figure}[t]
\epsscale{1}
\plotone{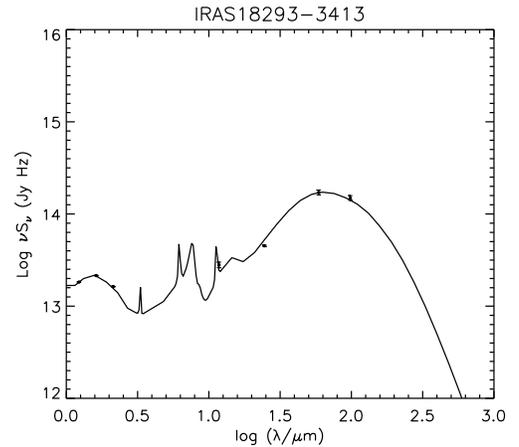}
\caption{
The IR SED of IRAS 18293-3413 fitted with the starburst models of Efstathiou
{\it et al.} (2000) and the cirrus model of Efstathiou and Rowan-Robinson
(2003). No AGN contribution was needed to explain the SED.}
\end{figure}

After correcting for a Galactic extinction of A$_{\rm V}$ = 0.47 (Schlegel et al. 1998) the absolute magnitude of the
SN at the time of the discovery becomes M$_{\rm Ks}$ = $-15.85$ (before correcting for the host galaxy extinction). In Fig.~3,
we compare the photometry of the SN with template K-band light curves for ordinary and slowly-declining CCSNe from MM01.
These templates comprise 2 and 3 linear components, respectively, and are based on the light curves of 11 (type II and Ib/c)
and two (the type IIL SN 1979C and the type IIn SN 1998S) CCSNe. The slow decliners are significantly more luminous than the 
ordinary events, even at early times, therefore requiring a larger host galaxy extinction to match the observations.
Assuming that the SN was discovered at the K-band maximum light, we need to dim the ordinary and slowly declining templates by 
A$_{\rm K}$ = 2.75 and 4.15 magnitudes, respectively, to match the observations. To estimate the lowest plausible extinction 
towards the SN we sought the latest possible SN epoch of discovery to match the templates. The time interval between the reference 
and discovery images was $\sim$130 days and the template light curves of MM01 have rise times of $\sim$25 days to the K-band
maximum light. The latest SN epoch of discovery assuming that the SN exploded soon after the observation of the reference image
is therefore about 100 days past the K-band maximum light. We found that the ordinary and slowly-declining templates can match 
the observations at this epoch if A$_{\rm K}$ = 0.55 and 2.50 magnitudes, respectively. Adopting the extinction law of Rieke 
\& Lebofsky (1985) the range of likely host galaxy extinction towards \object{SN 2004ip} is therefore between about 5 and 40 magnitudes in 
A$_{\rm V}$.

\begin{figure}[t]
\epsscale{1.15}
\plottwo{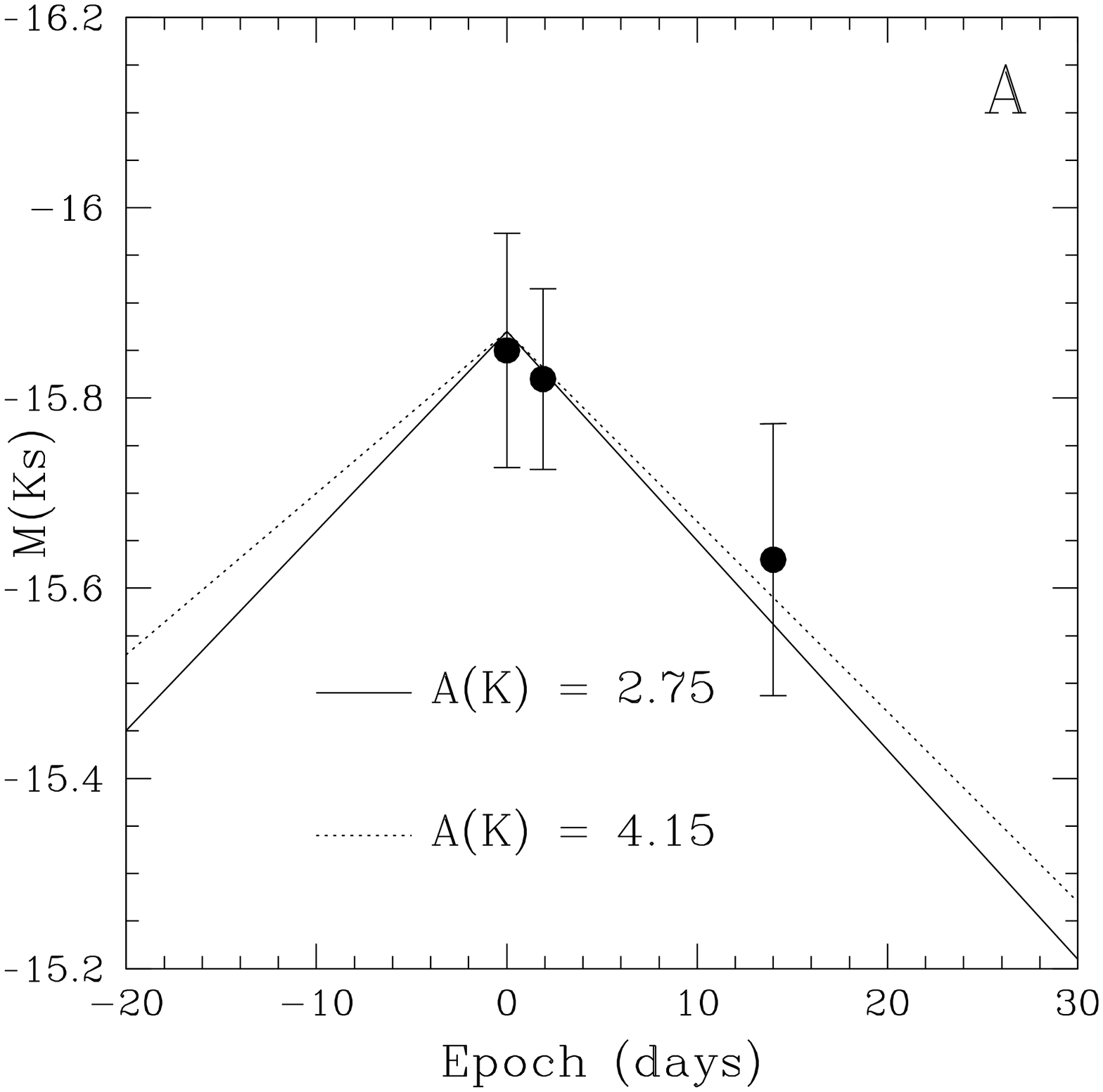}{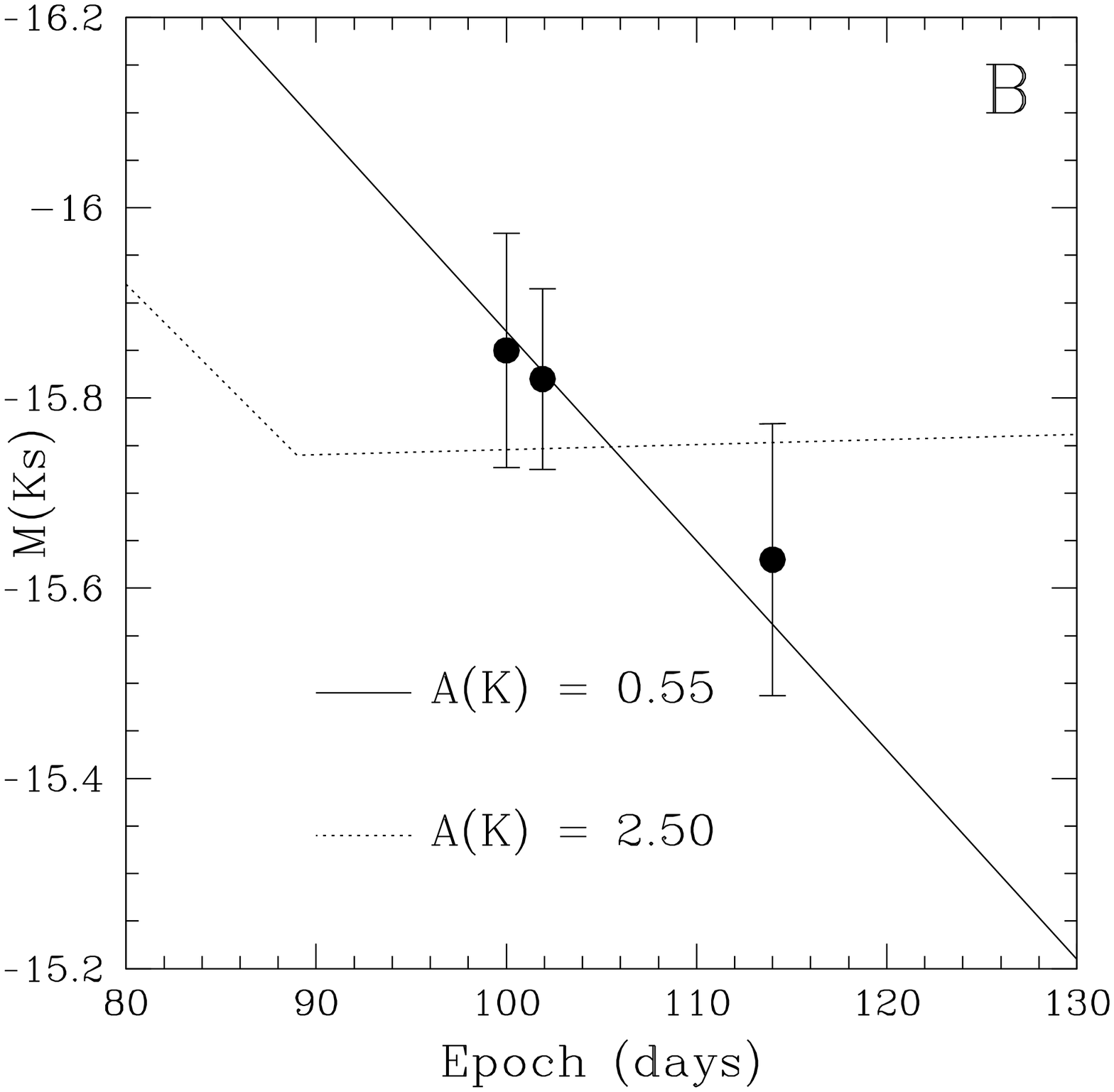}
\caption{Absolute magnitudes of SN 2004ip compared with template K-band light curves (MM01)
for ordinary (solid line) and slowly-declining (dotted line) CCSNe. The SN epoch of discovery was assumed to be at
(K-band) maximum light (A), 100 days past the maximum light (B). The host galaxy extinctions used for
dimming the light curves are shown in the figures.}
\end{figure}

\section{Discussion}
The number of CCSNe discovered in the infrared (Van Buren et al. 1994b; Maiolino et al. 2002;
Mattila et al. 2005a,b) is still rather small. However, these few events can have an important impact for the SN
statistics when estimating the {\it complete} local CCSN rate including also the SNe
in the optically obscured parts of the galaxies. Detecting CCSNe directly in starburst
galaxies over a range of IR luminosities (L$_{\rm IR}$ = 10$^{10}$--10$^{12}$ L$_{\odot}$) is also vital
for interpreting the results of the high-z CCSN searches since a large fraction of the massive
star formation at high-$z$ took place in IR luminous galaxies.

\object{SN 2004ip} is the first reported supernova discovered using adaptive optics. This discovery demonstrates the
great potential of the current 8-meter class telescopes equipped with AO correction in discovering SNe
from the innermost nuclear regions of LIRGs and ULIRGs. In addition, the next generation wide-field
AO systems can be used for expanding near-IR SN searches to the high-z Universe. Near-IR-discovered SNe in nuclear starbursts are not only
important for SN rate estimates, they can also be used to estimate the amount and distribution of extinction within
their host galaxies, and follow-up observations can provide us with a better understanding of the behaviour of SNe within
the dusty, high-density starburst environment. The extinction of up to 40 magnitudes in A$_{\rm V}$ estimated for
\object{SN 2004ip} is consistent with those derived towards the radio SNRs in the nuclear (central 500 pc)
regions of M~82 (MM01). We note that Infrared Space Observatory (ISO) studies (Genzel et al. 1998) indicate much higher
extinctions of A$_{\rm V}$(mixed) $\sim$ 50-1000 towards the highly-excited mid-IR line-emitting innermost nuclear regions
in LIRGs and ULIRGs. However, \object{SN 2004ip} is located in a circumnuclear starburst at a 500 pc distance from
the K$_{\rm S}$-band nucleus where the conditions could be expected to be similar to the M~82 nuclear environment.

Many of the SNe exploding in the nuclear starburst environment can be expected to produce very luminous radio SNe as a result
of the interaction of the SN ejecta with a dense surrounding medium (Chevalier and Fransson 2001). In fact, the radio discovered
SN 2000ft (Colina et al. 2001; Alberdi et al. 2006) in a circumnuclear starburst ring (600 pc from the nucleus) of its LIRG host NGC 7469, showed a
luminosity comparable to the brightest known radio SNe. Radio follow-up observation of \object{SN 2004ip} and other SNe
discovered in LIRGs and ULIRGs in the future can therefore provide important information on their progenitor stars and on the
conditions in the surrounding circumstellar medium.

\acknowledgements
This work was supported by funds from the Participating Organisations of EURYI and
the EC Sixth Framework Programme. This paper is based on observations collected at the
European Southern Observatory, Paranal, Chile (ESO Programmes 072.D-0433 and 073.D-0406).
We thank the ESO staff for carrying-out the service observations and C. Alard and
R. Joseph for useful discussions.


\begin{thebibliography}{}

\bibitem{b1} Alard, C. \& Lupton, R.H. 1998, ApJ, 503, 325
\bibitem{b2} Alard, C. 2000, A\&AS, 144, 363
\bibitem[Alberdi et al.(2006)]{2006ApJ...638..938A} Alberdi, A., Colina, 
L., Torrelles, J.~M., Panagia, N., Wilson, A.~S., \& Garrington, S.~T.\ 
2006, ApJ, 638, 938
\bibitem[]{342} Cappellaro, E., Evans, R., Turatto, M. 1999, A\&A, 351, 459
\bibitem[Cappellaro et al.(2005)]{2005A&A...430...83C} Cappellaro, E., et 
al.\ 2005, A\&A, 430, 83 
\bibitem[Chevalier \& Fransson(2001)]{2001ApJ...558L..27C} Chevalier, 
R.~A., \& Fransson, C.\ 2001, ApJL, 558, L27 
\bibitem[Colina et al.(2001)]{2001ApJ...553L..19C} Colina, L., Alberdi, A., 
Torrelles, J.~M., Panagia, N., \& Wilson, A.~S.\ 2001, ApJL, 553, L19 
\bibitem[Dahlen et al.(2004)]{2004ApJ...613..189D} Dahlen, T., et al.\ 
2004, ApJ, 613, 189
\bibitem[Efstathiou et al.(2000)]{2000MNRAS.313..734E} Efstathiou, A., 
Rowan-Robinson, M., \& Siebenmorgen, R.\ 2000, MNRAS, 313, 734
\bibitem[Efstathiou \& Rowan-Robinson(2003)]{2003MNRAS.343..322E} 
Efstathiou, A., \& Rowan-Robinson, M.\ 2003, MNRAS, 343, 322
\bibitem[]{355} Genzel, R., {\it et al.} 1998, ApJ, 498, 579
\bibitem[Grossan et al.(1999)]{1999AJ....118..705G} Grossan, B., Spillar, 
E., Tripp, R., Pirzkal, N., Sutin, B.~M., Johnson, P., \& Barnaby, D.\ 
1999, AJ, 118, 705 
\bibitem[Hachisu \& Kato(2006)]{2006ApJS..167...59H} Hachisu, I., \& Kato, 
M.\ 2006, ApJS, 167, 59
\bibitem[Lenzen et al.(2003)]{2003SPIE.4841..944L} Lenzen, R., et al.\ 
2003, PROCSPIE, 4841, 944 
\bibitem[Lonsdale et al.(2006)]{2006ApJ...647..185L} Lonsdale, C.~J., 
Diamond, P.~J., Thrall, H., Smith, H.~E., \& Lonsdale, C.~J.\ 2006, ApJ, 
647, 185 
\bibitem{maio} Maiolino, R., Vanzi, L., Mannucci, F. et al. 2002, A\&A, 389, 84
\bibitem[Mannucci et al.(2003)]{2003A&A...401..519M} Mannucci, F., et al. 
2003, A\&A, 401, 519
\bibitem{mattila} Mattila S. \& Meikle W.P.S. 2001, MNRAS, 324, 325
\bibitem[Mattila et al.(2004)]{2004NewAR..48..595M} Mattila, S., Meikle, 
W.~P.~S., \& Greimel R. 2004, New Astronomy Review, 48, 595
\bibitem[Mattila et al.(2005)]{2005IAUC.8473....1M} Mattila, S., et al. 
2005a, IAUC, 8473
\bibitem[Mattila et al.(2005)]{2005IAUC.8474....1M} Mattila, S., Greimel, 
R., Gerardy, C., \& Meikle, W.~P.~S 2005b, IAUC, 8474 
\bibitem[Mattila et al.(2007)]{} Mattila, S., {\it et al.} 2007, CBET, 858
\bibitem[Parra et al.(2006)]{2006astro.ph.12248P} Parra, R., et al. 2006, astro-ph/0612248
\bibitem[Risaliti et al.(2000)]{2000A&A...357...13R} Risaliti, G., Gilli, 
R., Maiolino, R., \& Salvati, M. 2000, A\&A, 357, 13 
\bibitem[Rousset et al.(2003)]{2003SPIE.4839..140R} Rousset, G., et al.\ 
2003 PROCSPIE, 4839, 140
\bibitem[Schlegel et al.(1998)]{1998ApJ...500..525S} Schlegel, D.~J., 
Finkbeiner, D.~P., \& Davis, M.\ 1998, ApJ, 500, 525 
\bibitem[Shafter(1997)]{1997ApJ...487..226S} Shafter, A.~W.\ 1997, ApJ, 
487, 226 
\bibitem[Strauss et al.(1992)]{1992ApJS...83...29S} Strauss, M.~A., Huchra, 
J.~P., Davis, M., Yahil, A., Fisher, K.~B., \& Tonry, J.\ 1992, ApJS, 83, 
29 
\bibitem[Sullivan et al.(2006)]{2006ApJ...648..868S} Sullivan, M., et al.\ 
2006, ApJ, 648, 868
\bibitem{b37} Van Buren, D. \& Greenhouse, M.A. 1994a, ApJ, 431, 640
\bibitem{b38} Van Buren, D., Jarrett, T., Terebey, S., Beichman, C.,
 Shure, M., Kaminski, C. 1994b, IAUCIRC, 5960, 2
\bibitem[Veilleux et al.(1995)]{1995ApJS...98..171V} Veilleux, S., Kim, 
D.-C., Sanders, D.~B., Mazzarella, J.~M., \& Soifer, B.~T.\ 1995, ApJS, 
98, 171
\end{thebibliography}
\end{document}